# Diamond molecular balance: Revolutionizing high-resolution mass spectrometry from MDa to TDa at room temperature

**Authors:** Donggeun Lee[1,+], Seung-Woo Jeon[1,+], Chang-Hwan Yi[2,+], Yang-Hee Kim[3], Yeeun Choi[1,4], Sang-Hun Lee[5], Jinwoong Cha[6], Seung-Bo Shim[6], Junho Suh[7], Il-Young Kim[1], Dongyeon Daniel Kang[1], Hojoong Jung[1], Cherlhyun Jeong[8,9], Jae-pyoung Ahn[3], Hee Chul Park[10,*], Sang-Wook Han[1,4,11,*], Chulki Kim[1,*]

[1]Center for Quantum Technology, Korea Institute of Science and Technology (KIST), Seoul 02792, Republic of Korea

[2]Center for Theoretical Physics of Complex Systems, Institute for Basic Science (IBS), Daejeon 34126, Republic of Korea

[3]Advanced Analysis and Data Center, Korea Institute of Science and Technology, Seoul 02792, South Korea

[4]KU-KIST Graduate School of Converging Science and Technology, Korea University, Seoul 02841, Republic of Korea

[5]Department of Optical Engineering, Kumoh National Institute of Technology, Gumi, Gyoungbuk 39235, Republic of Korea

[6]Quantum Technology Institute, Korea Research Institute of Standards and Science, Daejeon 34113, Republic of Korea

[7]Department of Physics, Pohang University of Science and Technology (POSTECH), Pohang 37673, Republic of Korea

[8]Chemical and Biological Integrative Research Center, Korea Institute of Science and Technology, Seoul 02792, Republic of Korea

[9]Division of Bio-Medical Science & Technology, University of Science and Technology (UST), Seoul, Republic of Korea

[10]Department of Physics, Pukyong National University, Busan 48513, Republic of Korea.

[11]Division of Quantum Information, KIST School, Korea University of Science and Technology, Seoul 02792, Republic of Korea

[+] These authors contributed equally: Donggeun Lee, Seung-Woo Jeon, Chang-Hwan Yi

[*] These authors jointly supervised this work: Chulki Kim, Sang-Wook Han, Hee Chul Park.

*email address:chulki.kim@kist.re.kr, swhan@kist.re.kr, hcpark@pknu.ac.kr

**Abstract:**

The significance of mass spectrometry lies in its unparalleled ability to accurately identify and quantify molecules in complex samples, providing invaluable insights into molecular structures and interactions. Here, we leverage diamond nanostructures as highly sensitive mass sensors by utilizing a self-excitation mechanism under an electron beam in a conventional scanning electron microscope (SEM). The diamond molecular balance (DMB) exhibits an exceptional mass resolution of 0.36 MDa, based on its outstanding mechanical quality factor and frequency stability, along with an extensive dynamic range from MDa to TDa. This positions the DMB at the forefront of molecular balances operating at room temperature. Notably, the DMB demonstrates its ability to measure the mass of a single bacteriophage T4 by precisely locating the analyte on the device. These findings highlight the groundbreaking potential of the DMB as a revolutionary tool for mass spectrometry at room temperature.

**Introduction**

Systematic analysis on molecules in a tissue or cell continues to be pursued with the development of mass spectrometry at the heart of proteomics. The task to identify and probe diverse molecules is quite challenging because of the inherent complexity and structural diversity of molecular components. Conventional mass spectrometry (MS) can identify species of molecular analytes in the range of 1 to 100 GDa by using soft ionization, electromagnetic fields to manipulate ions, and ensemble averaging of mass-to-charge ratios[1,2]. In recent years, the limitations of conventional mass spectrometry in terms of mass resolution and dynamic range have prompted the exploration of alternative approaches for mass analysis.

Unprecedented potential of nanoelectromechanical system (NEMS) for mass sensing have been explored by many research groups[3-13]. Nanomechanical resonators measure the mass of individual particles accreting on their surface. They are weighing the inertial mass of a target analyte in any charged state (ionized or neutral). Some of them started with measuring Au nanoparticle, IgM antibody complex[3], then moved on to Tantalum nanocluster[5,6], bacteriophage T5[7], and later measured the dry mass of bacteria[9]. Mass resolution with NEMS was attained down to yoctogram scale – a few number of atoms[4]. And they cover larger detection range, up to GDa ranges than the conventional MS[14]. This large dynamic range in measurement naturally makes them attractive to the mass spectrometry community since analyzing large mass analytes such as viruses and diverse biomarkers for various diseases as well as synthetic nanoparticles for nanomedicine is highly demanded[7].

Two characteristics of NEMS devices – namely, their minuscule mass and electromechanical degree of freedom – play critical roles in typical NEMS applications. These ingredients are apparent strengths in mass sensing tasks where the resonant frequency variation of NEMS reflects the mass of adsorbed analytes. NEMS-based spectrometer can be engineered for specific target mass ranges by adjusting its dimensions, as the frequency-to-mass relationship scales with the resonator's characteristic dimension, $d^{-4}$. This can be further enhanced by a proper choice of material.

Diamond is the hardest material on earth and famous for its durability. With recent improvements in fabrication technology, its limitations of poor deformability and relatively high brittleness can be lifted at the nanoscale[15,16]. Furthermore, their outstanding mechanical property, possibly resulting in high quality factor, is found to be insensitive to environmental temperature[17], although its maximum achievable elastic strain and strength are determined primarily by surface conditions of the given structure[18]. These remarkable characteristics of diamond and fabrication technologies at nanoscale are about to open a new avenue of diamond NEMS.

In this work, as a proof-of-concept toward room temperature mass spectrometry at a molecular level, we harness the diamond nanostructure as an ultra-sensitive mass sensor, to be coined diamond molecular balance (DMB). We investigated its structural properties and operating mechanism by conducting experiments in transmission electron microscope (TEM, Tecnai F20 G2, FEI) and scanning electron microscope (SEM, Quanta 3D FEG, and G4 HX, FEI) with computational support. Calibrated analytes across a wide mass range, from attograms to picograms, were evaluated. The outstanding mass sensitivity, along with an exceptionally wide dynamic range, is attributed to its extremely high mechanical quality factor of 236,000, to be the best in the world with a nanoscale device operating at room temperature. The frequency stability of the DMB was assessed to determine its mass resolution resulting from the device alone. Finally, in a seminal validation of its capability, the DMB weighs a single bacteriophage T4 at room temperature by precisely locating the analyte on top of the DMB. The high device integration density and multiplexed measurement capability of the DMB will facilitate large cross-section of the capture area and high-throughput analysis, significantly enhancing its applicability in fields such as virology and molecular biology.

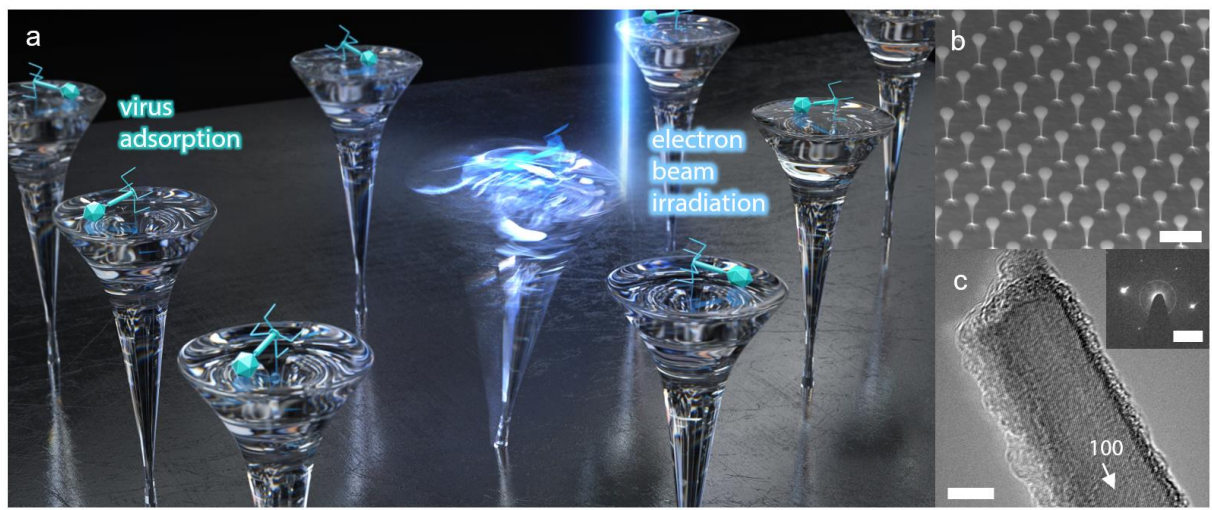

**Fig. 1. Diamond molecular balances. a** Illustration of diamond molecular balances with an incident electron beam. The resonant frequency of the DMB is detuned by the mass of the adsorbed analytes. **b** A scanning electron micrograph (SEM) image of DMBs (scale bar = 2 μm). **c** A tunneling electron micrograph (TEM) image of a DMB (scale bar = 5 nm), and (inset) its selected are electron diffraction pattern. The crystallinity of the nanostructure remains well preserved after fabrication processes. (scale bar = 4 $nm^{-1}$). The white arrow indicates (100) face of the diamond crystal structure.

**Diamond nanoscale structure and self-oscillation**

The DMB was fabricated by applying diamond fabrication techniques (see Supplementary Information 1 for details)[16]. We used a chemical vapor deposition (CVD) diamond substrate (2.0 × 2.0 × 0.5 $mm^3$) in single crystal electronic grade with ppb level of impurity (Element six, Electronic-Grade Single-Crystal Diamonds, ELSC20). A layer of a silicon nitride film was deposited and patterned by using electron beam lithography and a reactive ion etching process. By applying patterned silicon nitride disks as an etch mask, the diamond substrate was then selectively etched. A typical structure for DMB has dimensions of 600 nm in diameter at the top plateau and 1.5 μm in height. The diameter at the lower bottom of the structure varies within the range of 20 to 60 nm. The crystallinity of the structure after fabrication processes was confirmed by using TEM (Fig. 1c and Supplementary Information 2 for details).

The essence of the mass detection mechanism in our DMB lies in the self-oscillation induced by a primary electron beam[8,19]. Operating in a spot mode, a focused electron beam can be accurately positioned in a scanning electron microscope. As primary electrons interact with the diamond nanostructure, secondary electron emission induces positive charging. These localized charges prompt electrostatic interactions with the environment, resulting in nanostructure bending. Under given charge relaxation conditions, the diamond nanostructure undergoes self-oscillation through cyclic charging and discharging processes.

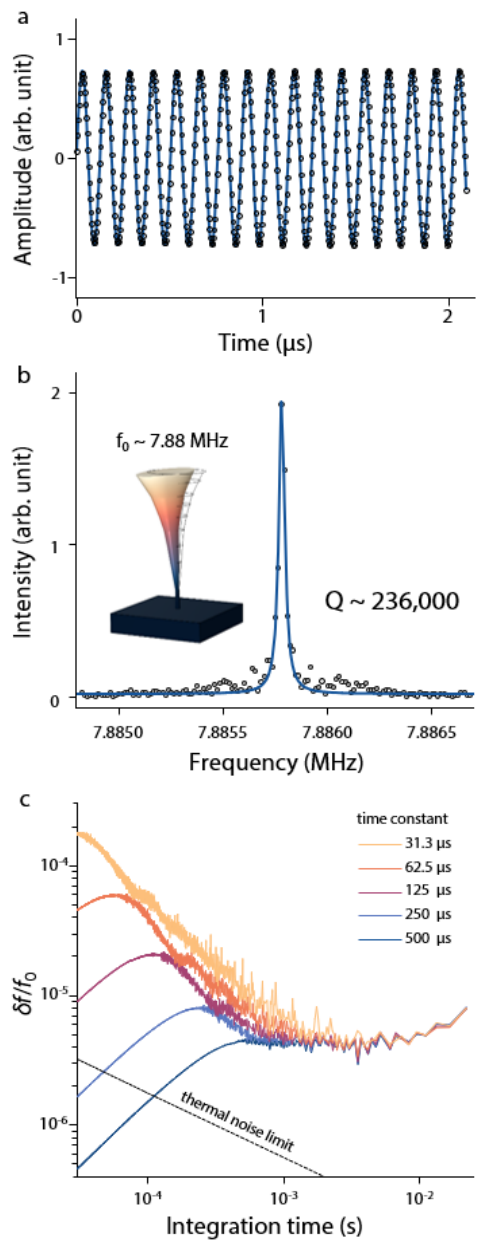

**Fig. 2. Self-oscillation of DMB and Fast-Fourier transform analysis. a** Time-dependent secondary electron intensity resulting from the self-oscillation of the DMB. Secondary electron intensity is periodically modulated by the mechanical motion of the self-oscillating DMB, and the time-dependent intensity signal is well traced by a sinusoidal function. **b** Fast-Fourier transform (FFT) result. From the Lorentzian curve fitting, a quality factor of 236,000 was extracted. (inset) Finite element simulation of a resonant mechanical mode of the DMB. The obtained resonance frequency closely matches the finite element calculation result of 7.88 MHz. **c** Frequency stability of the DMB under different time constant conditions. The dashed line indicates the thermal noise limit.

With its self-oscillation of the DMB, the amplitude of the secondary electron current is periodically modulated as seen in Fig. 2a (see Supplementary Information 3 for details). The secondary electron signal inherently contains the information on the mechanical oscillation at nanoscale as the primary electron beam scans over the DMB during its oscillation. We note the outstanding quality factor (~ 236,000) of this system in the frequency domain at room temperature (Fig. 2b). This agrees with the observation that the mechanical quality factor in diamond-based nanostructures is relatively insensitive to environmental temperature[17]. The frequency stability $\delta f/f_0$, based on the Allan deviation as shown in Fig. 2c, is assessed to be $3 \times 10^{-6}$, corresponding to the mass resolution of 0.36 MDa at room temperature (see Supplementary Information 14).

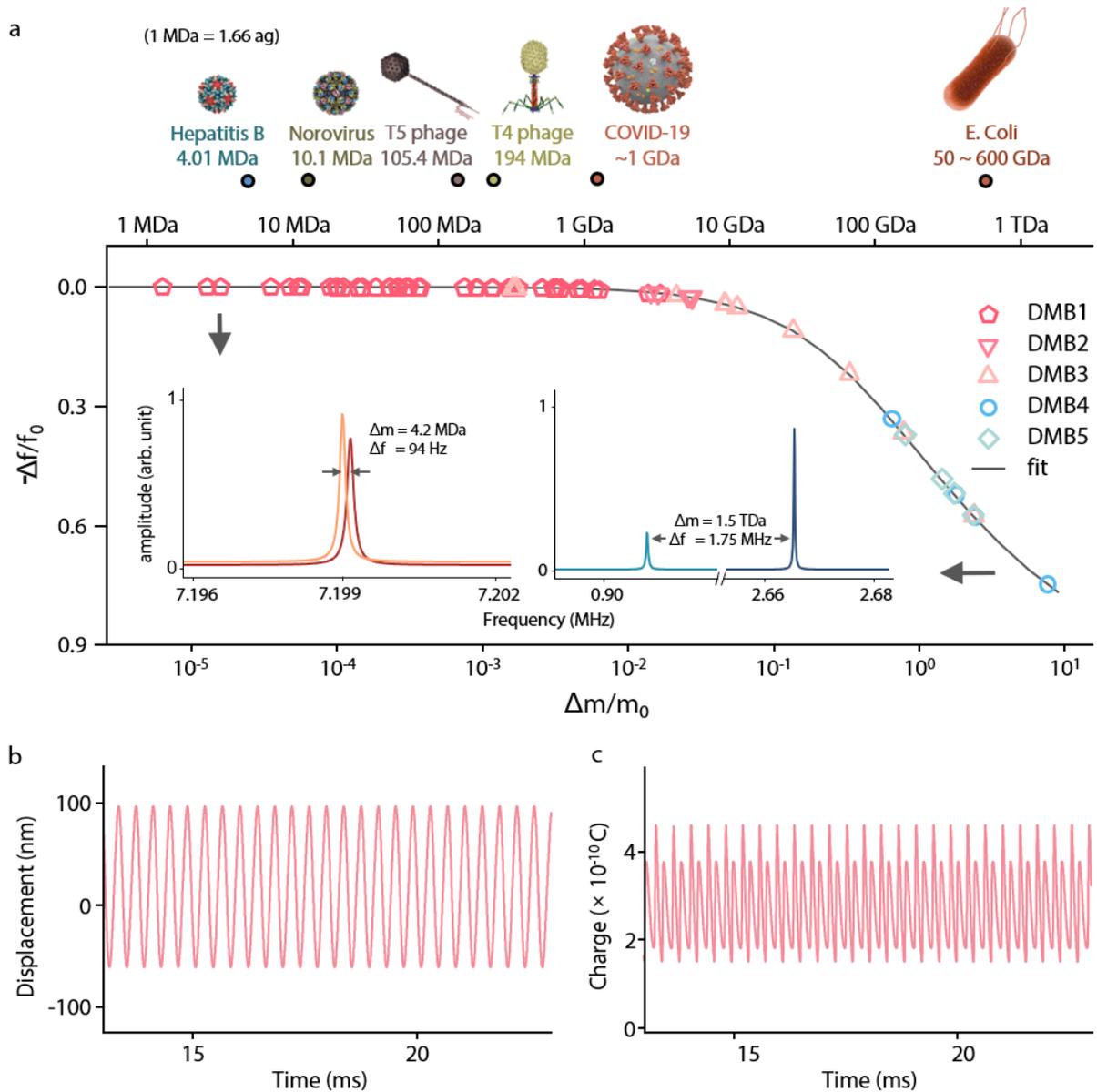

**Fig. 3. Mass detection range of DMB and a theoretical modeling. a** (upper) Molecular weights of various viruses and E. Coli. (bottom) Normalized frequency detunings as a function of normalized weights. The measured frequency detunings from five different DMBs with calibrated mass loading in the range of MDa to TDa (pentagon, inverted triangle, triangle, circle, and diamond, see Supplementary Fig. 18) were traced by numerically obtained results based on the theoretical model (solid line, see Supplementary Information 6 for details). (inset) Frequency shifted responses with the mass loadings of 4.2 MDa and 1.5 TDa. **b** Time-dependent displacement of DMB. **c** Time-dependent charge state of DMB.

**Ultra-wide mass detection range**

To explore the mass detection range of DMB, we loaded calibrated mass target by depositing carbon and platinum composites in different volumes on top of the DMB (see Supplementary Information 4 for details). The deposited weight was carefully calculated considering the obtained volumes, ratios of constituent elements, and their density information. The obtained frequency detunings are summarized according to loaded weights in Fig. 3a. Surprisingly, our DMB demonstrated quantitative mass detection in the range from 1.8 ag to 2.5 pg (1.1 MDa to 1.6 TDa). To date, it measures the widest range (6 orders of magnitude) of masses among the implemented methods for mass detection. The heaviest analyte was nearly 25 times heavier than the effective mass of the DMB itself. This exceptionally wide mass detection range encompasses various biological objects, from Hepatitis B virus to E. Coli, that cause transmissible infections (Fig. 3a). We note that the spring constant of the DMB stays almost unchanged with the standard deviation of 0.3 % over the course of 50 successive mass loading events (see Supplementary Fig. 21).

The electromechanical behavior of the DMB can be modeled by a driven-damped harmonic oscillator coupled to charge dynamics induced by an incident electron beam (see Supplemental Information 6 for details). The obtained frequency response to mass loading closely matches the numerically obtained results based on the proposed theoretical model. These calculations utilized experimental parameter values, including charge relaxation time, resonance frequency and the quality factor of the DMBs. From the theoretical model, validated by the experimental results, the effective mass of our system as well as frequency detunings at different mass loadings were extracted. We further investigated the time-dependent displacement and charge state of the DMB using parameters obtained from experimental results, as depicted in Fig. 3b and c. The mechanical motion of the DMB manifests as a limit cycle in phase space (Supplementary Fig. 22), exhibiting behavior that closely resembles that of a harmonic oscillator.

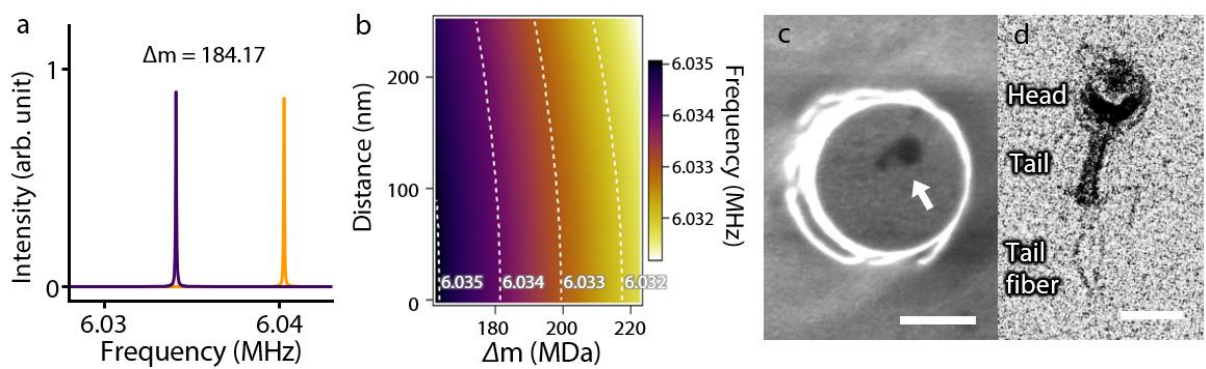

**Fig. 4. Mass measurement on a single bacteriophage T4**. **a** Detuned frequency response of a DMB before (yellow) and after (purple) loading a single bacteriophage T4. **b** Corrections on frequency detunings as a function of varying positions along the moving direction, the distance from the axis perpendicular to the moving direction, and masses of analytes on the plateau of the DMB. The dashed lines on the plot represent equal values of correction on the frequency detuning. **c** Scanning electron micrograph image (top view) of a single bacteriophage T4 on top of the DMB (scale bar = 300 nm). **d** Transmission electron micrograph image of a bacteriophage T4, revealing detailed body structure including tail fibers (scale bar = 100 nm).

**Mass measurement on a single bacteriophage T4**

As a notable example of mass measurement of biological analytes at room temperature, we conducted measurements on a single bacteriophage T4. Bacteriophage T4 (*Escherichia virus* T4) is a relatively large, double-stranded DNA virus that infects Escherichia coli bacteria. It has a distinctive head-and-tail structure, and a mature phage typically has dimensions of 90 nm in width and 200 nm in length with a weight of approximately 194 MDa (0.322 fg). Metallic layers of Ti and Au, each with a thickness of 5 nm, were deposited on the DMBs to enhance the contrast of the analytes under the electron beam. We applied a calculated volume concentration of the bacteriophage T4 solution to facilitate the adsorption of analytes on the surface of the DMB (see Supplementary Information 8 for details). The adsorbed bacteriophage T4 virions were neatly positioned on the DMB, clearly distinguishable from the top plateau surface. Almost no residual substances, including salt and buffer materials, were observed in the surface elemental analysis, suggesting effective rinsing during sample preparation (see Supplementary Fig. 9). During the adsorption process, the structure of the analyte remained intact as evidenced by the TEM image, revealing detailed body structure of bacteriophage T4 including its tail fibers (Fig. 4d). It was demonstrated that the DMB can accurately measure the inertial mass of bacteriophage T4 virions (Fig. 4a). A distinct resonance peak shift, corresponding to a mass of 184 MDa, was observed in the frequency spectrum for a single bacteriophage T4 (Fig. 4a). To address the position dependence of the frequency response, we calculated corrections for frequency detunings as a function of varying positions along the moving direction, the distance from the axis perpendicular to the moving direction, and masses of analytes on the plateau of the DMB (see Fig. 4b and Supplementary Information 9 for details). A mass resolution of 4.07 MDa was achieved at room temperature, with pressure fluctuation and the positional information limited by the spatial resolution (approximately 10 nm) of the scanning electron microscope (see Supplementary Information 10 for details). As seen in the SEM image the analyte exhibits a finite contact area rather than a point contact. Additionally, the stiffness of the analyte results in a downward shift and broadening of the resonance peak[7,20,21]. However, finite element simulations have shown that the impact of these factors is negligible compared to other determinants of mass resolution (see Supplementary Information 10 for details). This could be attributed to the minimal displacement induced by the DMB. The mass resolution is primarily degraded by the pressure fluctuations[22]. This issue can be mitigated by adopting an in-situ analyte loading system that maintains an optimal vacuum level, thereby minimizing pressure fluctuations. The calculated mass from the frequency detuning shows slight discrepancy from typical estimates. The analyte is most likely to be an immature bacteriophage within the sample batch[23]. Measurements from other DMBs yielded slightly varying mass values, all near the typical mass of 194 MDa (see Supplementary Fig. 11). There have been instances where measurement was taken with two or more bacteriophage T4 virions simultaneously on a single DMB, resulting in mass values that are multiples of the typical mass.

**Discussion**

The DMB, as described above, marks a significant advancement in research capabilities for analyzing biological specimens. Operating at room temperature, it eliminates the need for low-temperature apparatus, streamlining experimental setups. This capability is particularly beneficial for studying molecular interactions in pathological conditions, enhancing our understanding of complex biological processes. Moreover, the impressive 6 orders of mass detection range from MDa to TDa with exceptional sensitivity demonstrated by the DMB represents a significant leap forward for next-generation mass spectrometers, addressing a critical gap in existing weighing technologies[7]. Notably, its ability to accurately weigh individual bacteriophage T4 virions underscores its potential for molecular-level mass spectrometry, positioning it as a transformative tool for probing molecular dynamics.

Adopting a top-down fabrication approach, the DMB achieves high device integration density without requiring supplementary circuitry for measurement, thus augmenting capture cross-section (see Supplementary Information 12 for details). The implementation of multiplexed measurement on highly integrated DMBs with controlled electron beam positioning, (see Supplementary Information 13 for details) holds promise for the development of high-throughput applications for intensive analysis. Finally, the combination of multiplexed measurement equipped with an in-situ analyte loading technique, such as electrospray ionization, presents a potential avenue for advanced mass spectrometry with an ultra-wide mass detection range at room temperature.